\documentclass[twocolumn,prb,showpacs,multicol,amsmath,amssymb]{revtex4}
\usepackage[dvips]{graphicx}
\usepackage{color}
\newcommand{\be}{\begin{equation}}
\newcommand{\ee}{\end{equation}}

\newcommand{\bea}{\begin{eqnarray}}
\newcommand{\eea}{\end{eqnarray}}
\newcommand{\bd}{\begin{displaymath}}
\newcommand{\ed}{\end{displaymath}}
\newcommand{\ba}{\begin{array}}
\newcommand{\ea}{\end{array}}
\newcommand{\bi}{\begin{itemize}}
\newcommand{\ei}{\end{itemize}}
\newcommand{\bc}{\begin{center}}
\newcommand{\ec}{\end{center}}
\newcommand{\bfl}{\begin{flushleft}}
\newcommand{\efl}{\end{flushleft}}
\newcommand{\bfr}{\begin{flushright}}
\newcommand{\efr}{\end{flushright}}

\def\6{\partial}

\def\={\!\!\!&=&\!\!\!}
\def\+{\!\!\!&&\!\!\!+~}
\def\-{\!\!\!&&\!\!\!-~}

\begin{document}
\date{\today}
\title{Magnetic impurity resonance states and symmetry of the superconducting order parameter in iron-based superconductors}

 \author{Alireza Akbari$^{1}$}
\author {Ilya Eremin$^{1,2}$}
\author {Peter Thalmeier$^{3}$}
\affiliation{ $^{1}$Max-Planck-Institut f\"ur Physik komplexer Systeme, N\"othnitzer Str.38,
01187 Dresden, Germany\\$^2$ Institute f\"ur Mathematische und Theoretische  Physik,
TU-Braunschweig,
D-38106 Braunschweig, Germany \\
$^{3}$Max-Planck-Institut f\"ur Chemische Physik fester Stoffe,
D-01187 Dresden, Germany}

 \begin{abstract}
We investigate the effect of magnetic impurities on the
local quasiparticle density of states (LDOS) in iron-based superconductors. Employing the two-orbital model
where 3$d$ electron and hole conduction bands are hybridizing with the localized
$f$-orbital of the impurity spin, we investigate how various symmetries of the superconducting gap and its
nodal structure influence the quasiparticle excitations and impurity bound states. We show that the bound states behave qualitatively different for each symmetry. Most importantly we find that the impurity-induced bound states can be used to identify the nodal structure of the extended $s$-wave symmetry ($S^{\pm}$) that is actively discussed in ferropnictides.
 \end{abstract}

\pacs{74.20.Rp, 74.25.Ha, 74.70.Tx, 74.20.-z  }

\maketitle


 \section{Introduction}

The problem of magnetic impurities in a superconductor has been extensively
discussed in the literature\cite{Shiba68,Matsumoto02,Sakai93,Satori92,Balatsky06}.
The magnetic impurity and its moment can  interact with the conduction
electrons of the metal in the normal or  superconducting state. In the former case this leads to
the Kondo effect and a resonance state at the Fermi level.
In the latter case it is well known that a single magnetic impurity doped into a superconductor produces
a localized bound state within the quasiparticle excitation gap\cite{Shiba68}. The spectrum is sensitive
to the symmetry of the order parameter and is therefore a powerful tool to probe the pairing symmetry.

The discovery of new Fe-based superconductors\cite{Kamihara08} with distinct multi-orbital
band structure \cite{Bruning08,Zhao08,Pourovskii08}
have opened a new horizon to high temperature superconductivity.
One of the most significant questions for these materials is the symmetry of the
superconducting gap and the underlying Cooper-pairing mechanism. The latter is believed to arise due to purely electronic mechanism and a variety of models have been
investigated with various weak-coupling approaches within random phase approximation RPA \cite{Kuroki08,Maier09,Schmalian09} and
renormalization group techniques \cite{Chubukov08,Zhai09}.
It was concluded that the fully gapped extended s-wave state with the $\pi$-shift of the gap between electron and
hole Fermi surface sheets is the most natural outcome of these theories. It is believed to be driven by the interband spin fluctuations at the antiferromagnetic
wave vector $(\pi,\pi)$ in folded Brillouin zone and it also competes with the spin density wave instability at the
same wave vector which leads to the columnar or striped AF state for low doping.

However, despite intensive
experimental efforts, the pairing symmetry of this new class of
superconducting materials is not completely settled.
Some experimental groups have reported the fully gapped
behavior \cite{Lue09,Hashimoto09,Malone09,Khasanov08,Zhang01}, but some measurements, in particular NMR relaxation and penetration depth suggest existence of gap nodes \cite{Nakai08,Shan08,Mu08}. From the theoretical side it has also been realized\cite{vavilov,Maier09} that the superconducting gap structure may be non-universal in ferropnictides due to the large intraband Coulomb repulsion. Its inclusion may force the superconducting gap to develop a node which crosses one of the Fermi surfaces. At the same time the symmetry of the superconducting gap will, however, still remain extended $s$-wave though  higher harmonics are acquired. Moreover, in some scenarios\cite{moreo,kuroki} the superconducting gap even changes from the extended $s$-wave towards either $d_{x^2-y^2}$- or $d_{xy}$-wave symmetries depending on the slight variation of parameters. Recently it has been found that isoelectronic substitution of As by P in BaFe$_2$(As$_{1-x}$P$_x$)$_2$ changes the gap structure in Fe pnictide compounds from nodeless to nodal \cite{Hashimoto09}. It seems that whereas electron- and hole doping leads to  a fully gapped state, isoelectronic doping (equivalent to chemical pressure) leads to the presence of line nodes in the gap function.
Therefore it is desirable to investigate the magnetic impurity effect in Fe pnictide compounds for different candidates of gap symmetries. The resulting characteristics of the  LDOS which is sensitive to the nodal structure may provide a clue to  distinguish between the various proposed gap symmetries.
In previous investigations for FeAs compounds the effect of nonmagnetic impurities on the quasiparticle spectrum in the $S^\pm$ state \cite{Matsumoto09,Zhang09} has been studied. Magnetic impurity effects  have  so far only been discussed for the single band model with d$_{x^2-y^2}$ order parameter \cite{Zhang01}, and for the two-band model for  a classical local moment\cite{Tsai09}. At the same time, it is known that the local density of state around the magnetic impurity can provide significant information on the local electronic structure in the unconventional superconductor\cite{hudson}. Note also that the influence of nonmagnetic \cite{Senga08,Onari09} and magnetic \cite{Li09} impurities on the reduction of superconducting transition temperature has been recently analyzed.

In this paper we investigate the effect of a single magnetic impurity on the local quasiparticle excitations around the impurity site. We use
a minimal two-band model for the electronic structure of Fe pnictides which leads to the
$\Gamma (0,0)$- centered hole and M $(\pi,\pi)$-centered electron pockets.
In section~\ref{sect:model} the Anderson model with a strong Hubbard  repulsion for the localized f-electron at
the impurity site, and a hybridization between conduction bands and localized  state will be introduced. We will
treat this model in the infinite U limit where a slave boson representation may be used similar to Ref.~\onlinecite{Zhang01}
where the Anderson impurity in the single band model with d$_{x^2-y^2}$ order parameter has been studied.
We then calculate the local density of states (LDOS) and discuss the signatures of possible Fe pnictide order parameter
symmetry in its spectral and spatial characteristics. This quantity is accessible in STM tunneling spectroscopy \cite{Balatsky06}.
The numerical results for the various cases will be discussed in section~\ref{sect:numerics}.
Finally in section~\ref{sect:summary} we give a summary  of our results and a conclusion.

\begin{figure}
\centerline{
  \includegraphics[angle=0,width=0.5\linewidth]{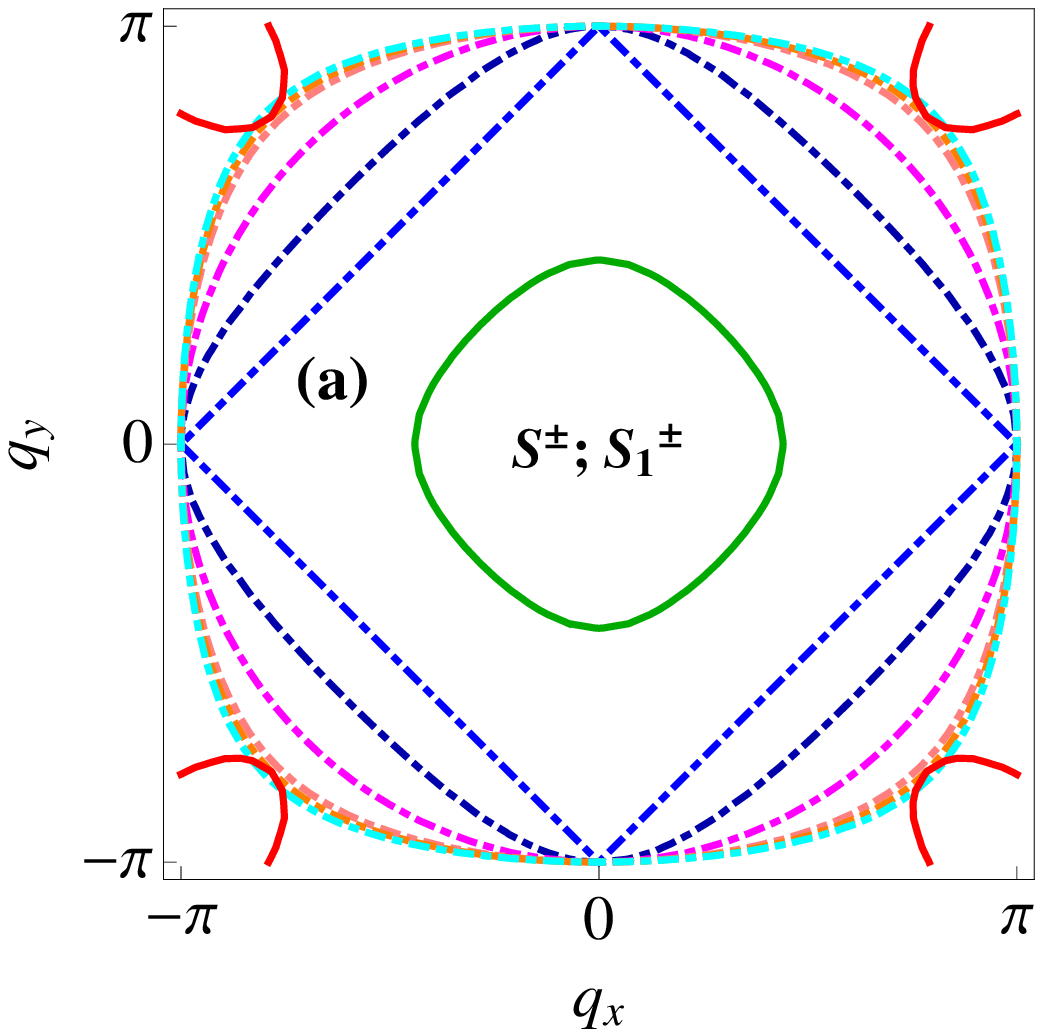}
\includegraphics[angle=0,width=0.5\linewidth]{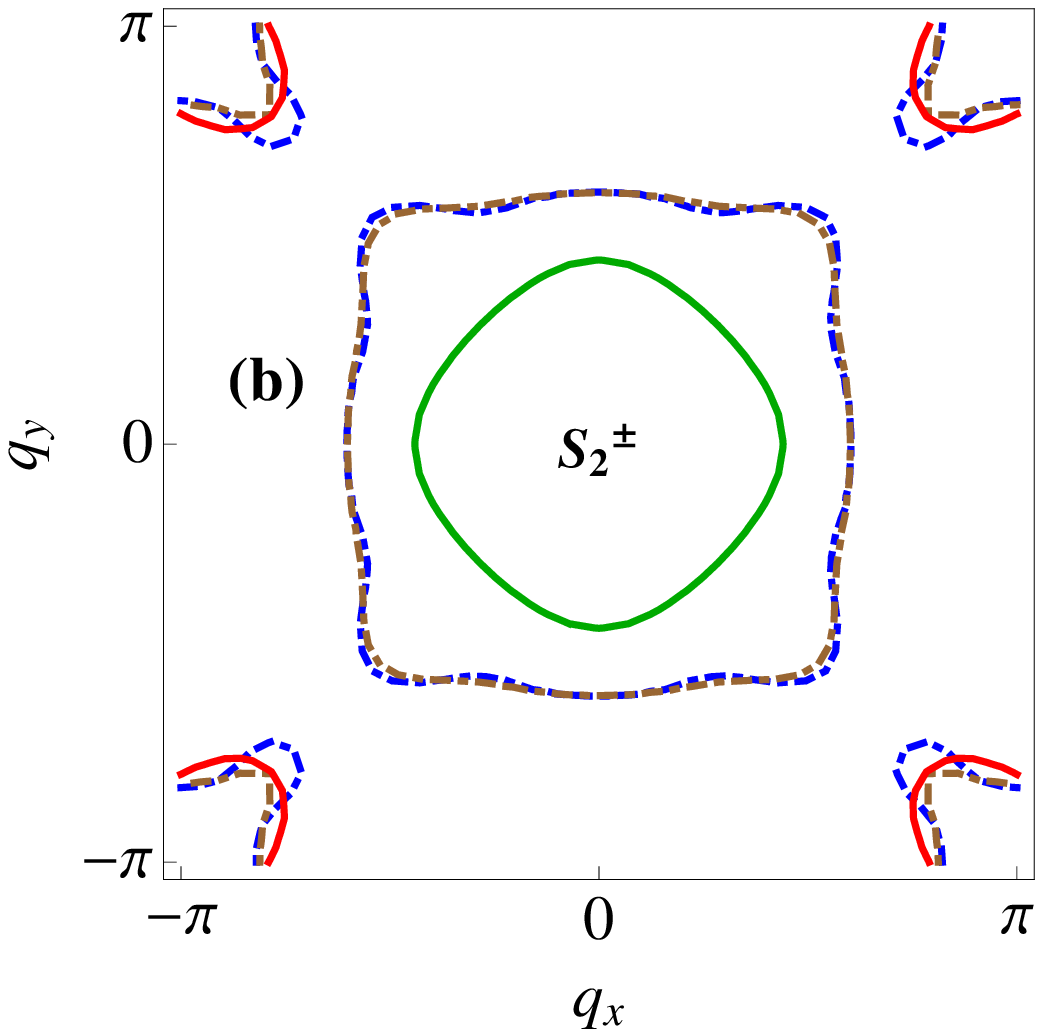}
}
\centerline{\includegraphics[angle=0,width=0.5\linewidth]{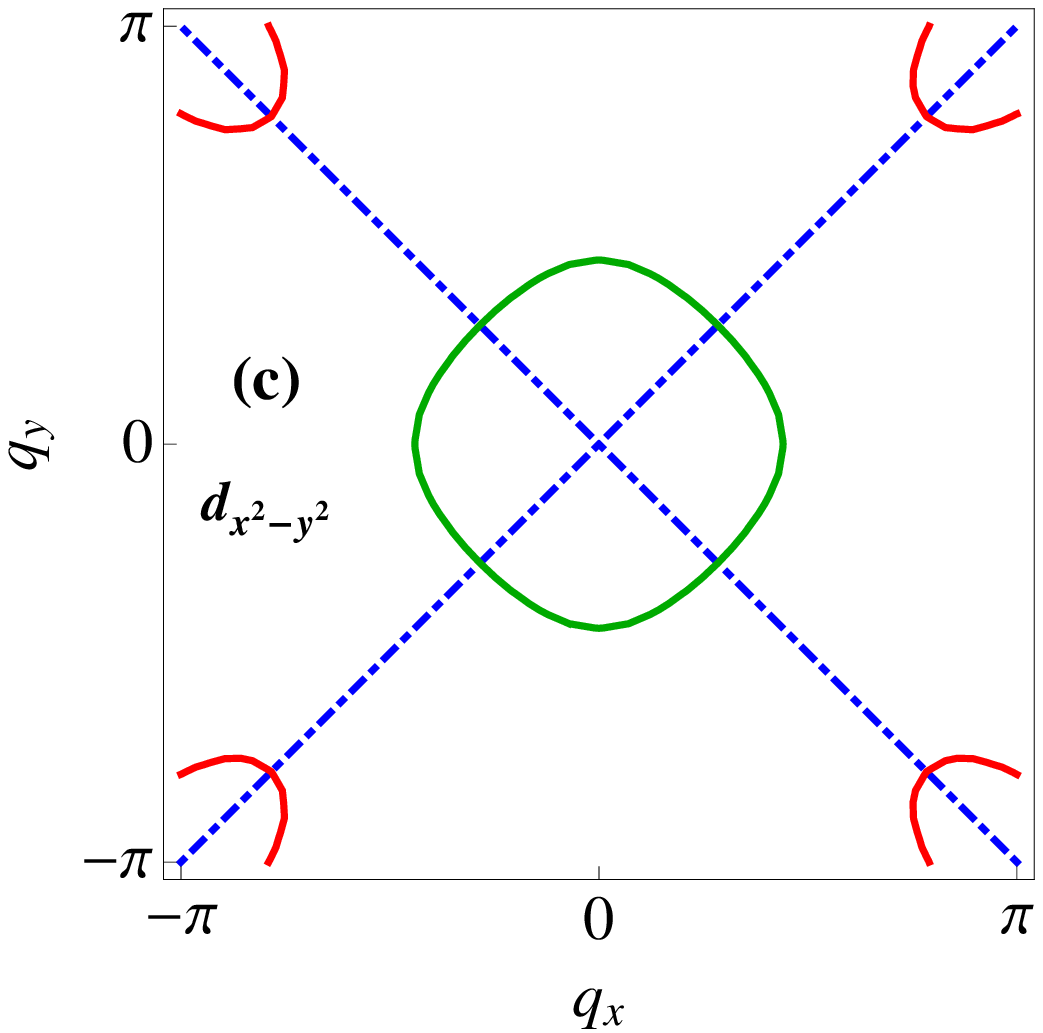}
\includegraphics[angle=0,width=0.5\linewidth]{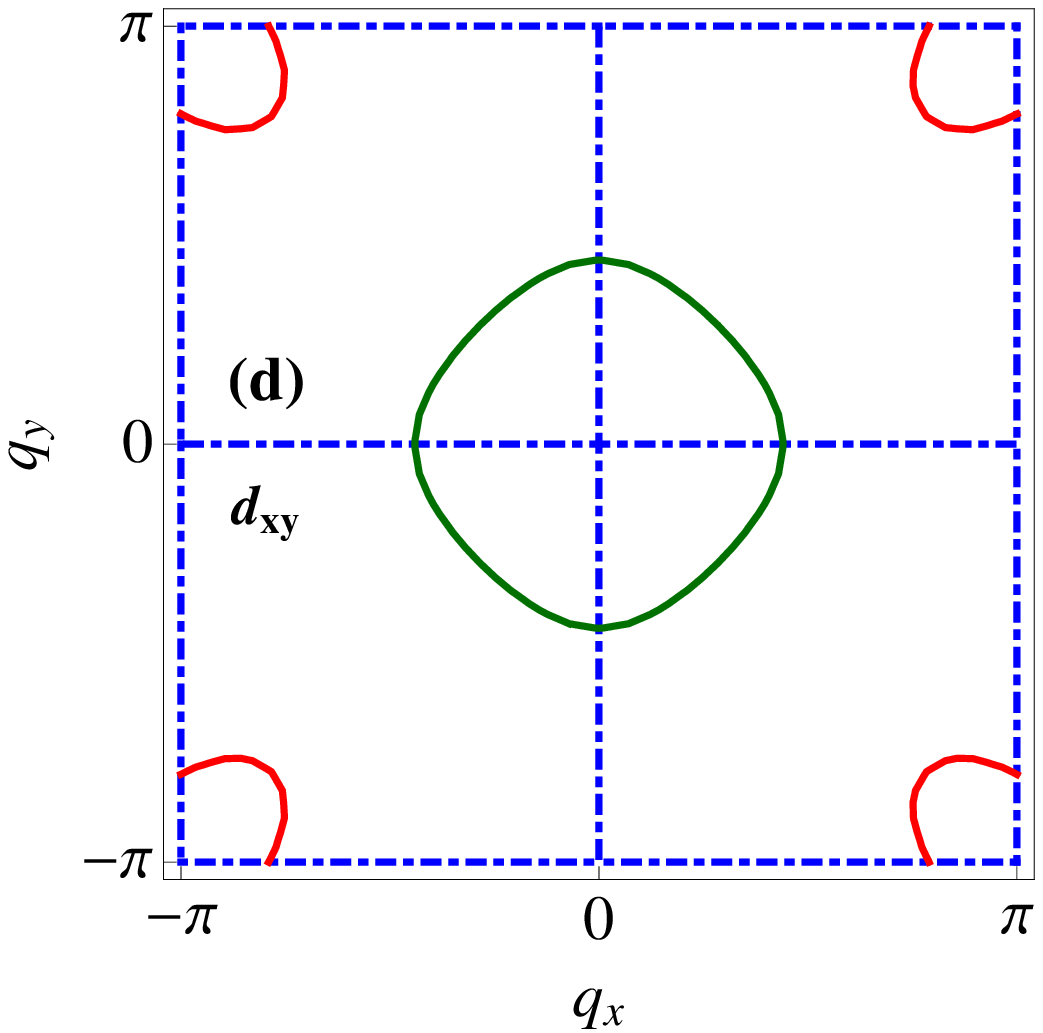}
}
\caption{(color online) Fermi surfaces of the two-band model (red and green thick curves)
with different  symmetry of the superconducting order parameters (nodal lines represented by the dashed lines).
(a) extended s-wave symmetry, starting from fully gapped
$\Delta_{S^\pm}(k_x,k_y)$ (or $\Delta_{S_1^\pm}(k_x,k_y)$ with $\alpha=0$).  With increase
of the higher harmonics,  $\alpha=1,3,6,7,8$ in $\Delta_{S_1^\pm}(k_x,k_y)$ this gap becomes
more anisotropic and finally has accidental nodes on the electronic pocket around the M-point;
(b) refers to the other extended  nodal s-wave gap symmetry  $\Delta_{S_2^\pm}(k_x,k_y)$ with two separate nodal lines: $\Delta_{S_2^\pm}(k_x,k_y)$ for $\alpha=1.2;\; \alpha'=0.15$ and $\alpha=1.17;\; \alpha'=0.08$);
(c)-(d) show $d_{x^2-y^2}$ and $d_{xy}$ gap symmetries with symmetry enforced nodes.
}
\label{figFS}
\end{figure}

\begin{figure}
\centerline{
\includegraphics[angle=0,width=0.5\linewidth]{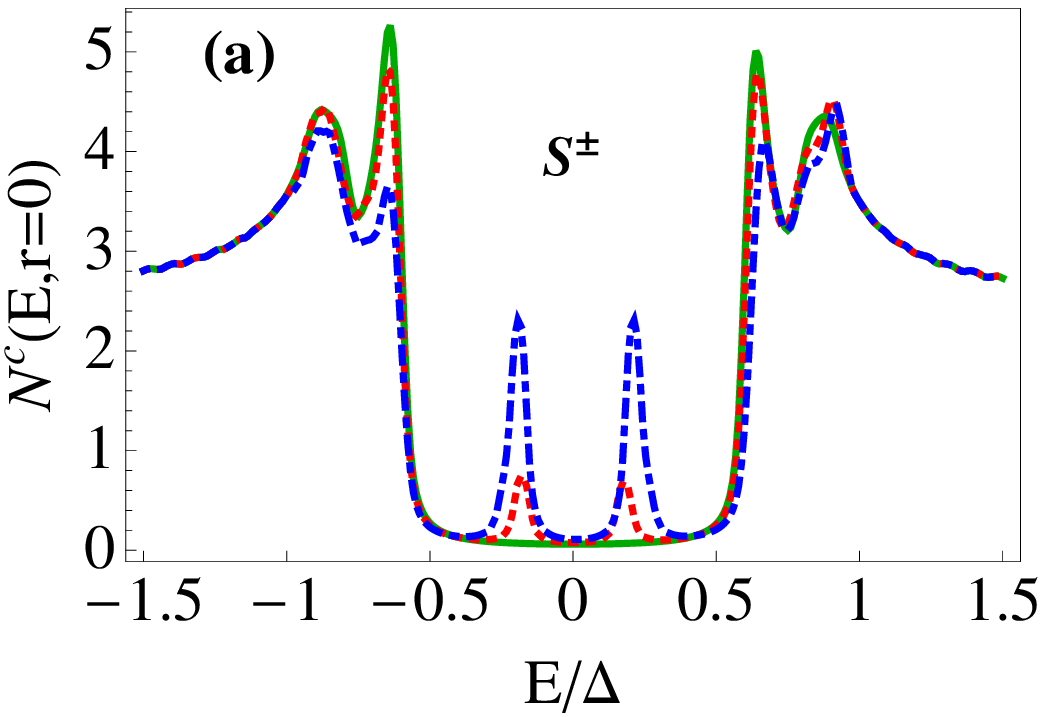}
\includegraphics[angle=0,width=0.5\linewidth]{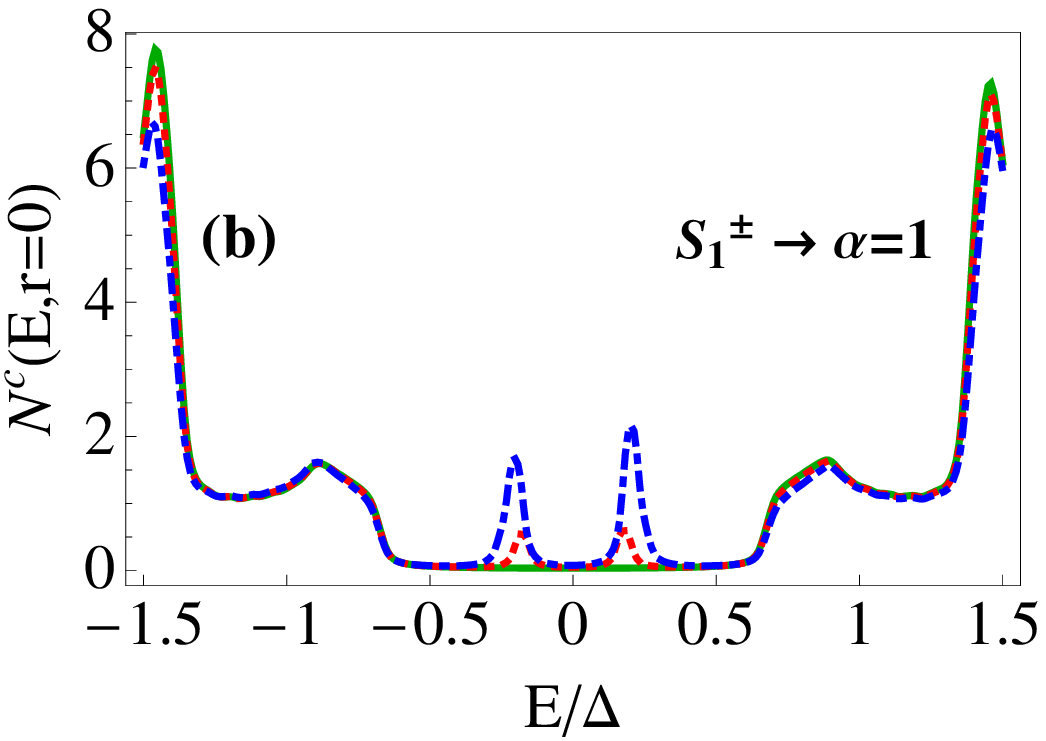}
}
\centerline{
\includegraphics[angle=0,width=0.5\linewidth]{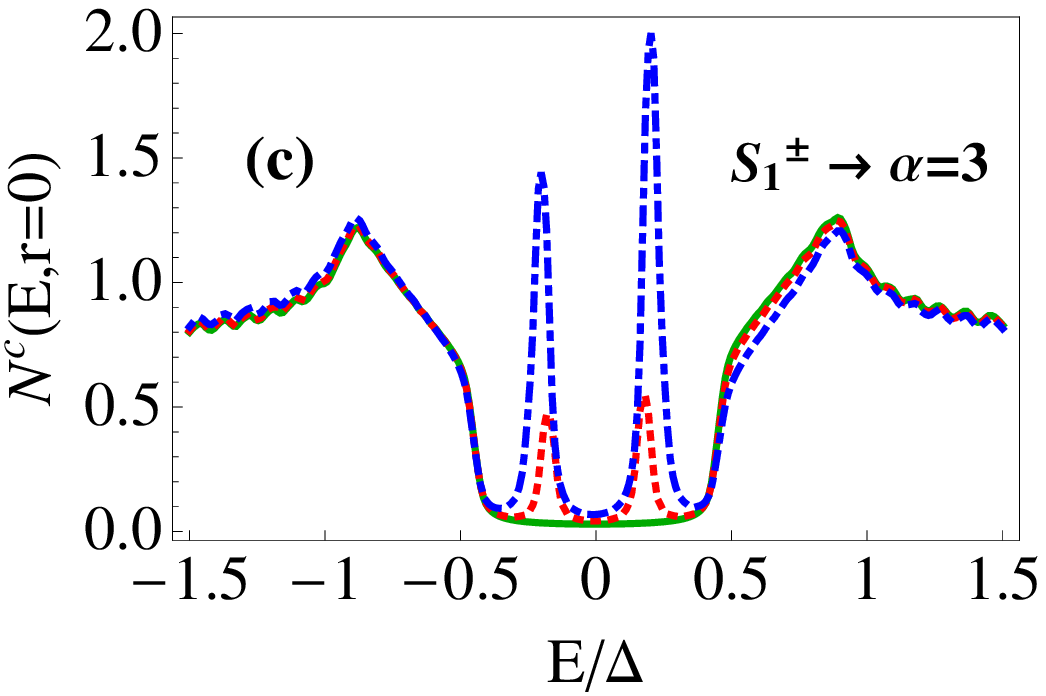}
\includegraphics[angle=0,width=0.5\linewidth]{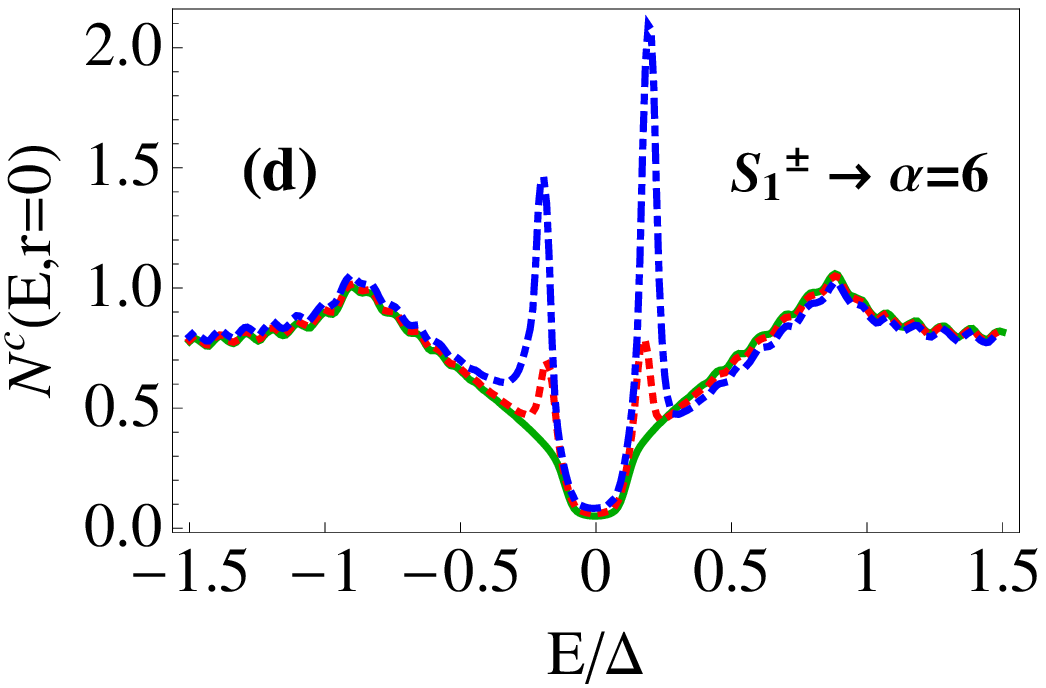}
}
\centerline{
\includegraphics[angle=0,width=0.5\linewidth]{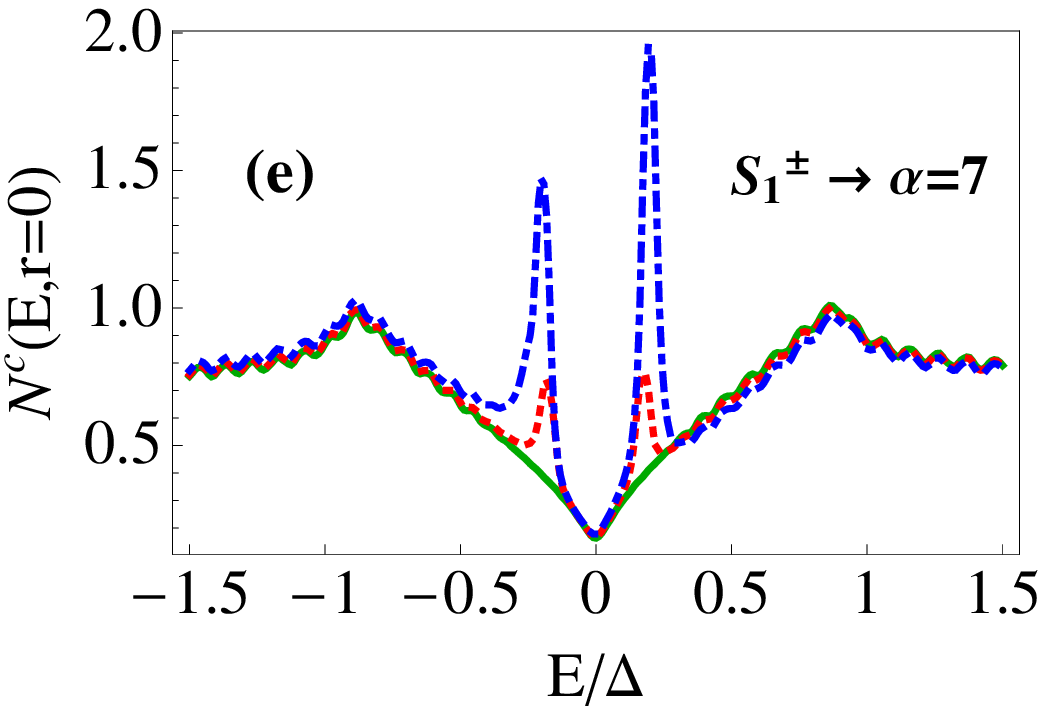}
\includegraphics[angle=0,width=0.5\linewidth]{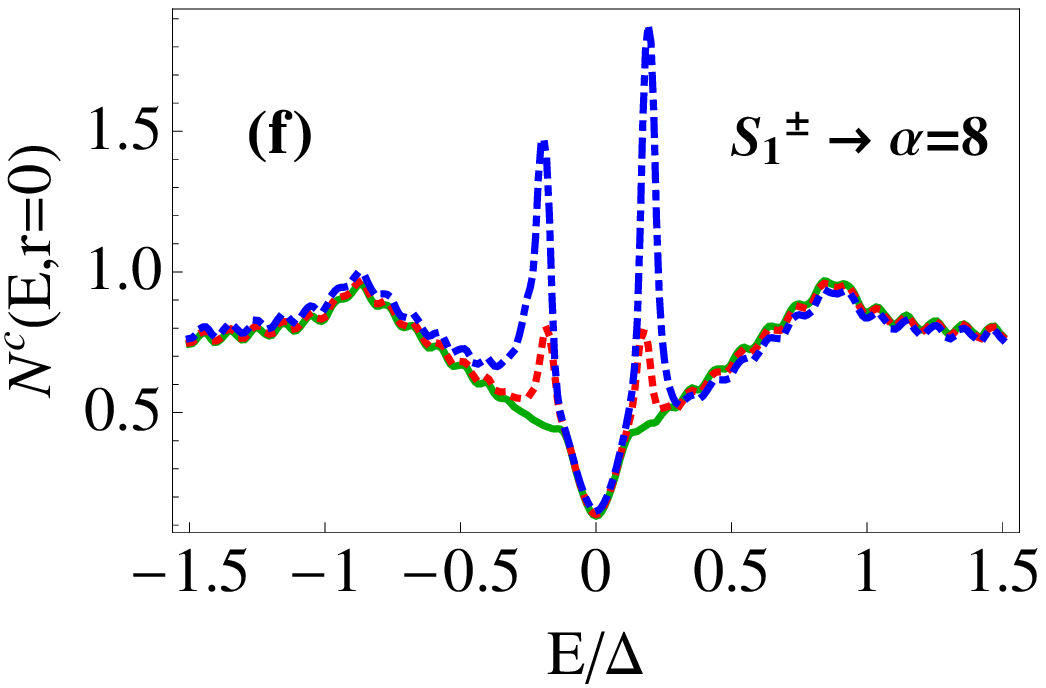}
}
\caption{(color online) Calculated local density of states (LDOS) for the superconducting regime with various superconducting order parameters: a) $S^\pm$: $\Delta_{S^\pm}(k_x,k_y)$ and (b)-(f) $S_1^\pm$:
 $\Delta_{S_1^\pm}(k_x,k_y)$ (with $\alpha=1,3,6,7,8$ respectively); for $\epsilon_f=-\Delta_0/3$,  $V_{1}=V_{2}=0$ (green),
 $V_{1}=V_{2}=0.5\Delta_0$ (red),
 and  $V_{1}=V_{2}=\Delta_0$ (blue).}
\label{Pspm1}
\end{figure}


\section{Theoretical Model}
\label{sect:model}

According to the band structure calculations\cite{LDA} as well as numerous ARPES results\cite{arpes} the Fermi surface topology of  iron-based superconductors consists of the small size circular hole and elliptic electron Fermi surface pockets centered around the $\Gamma-$ and $(\pi,\pi)$-points of the folded BZ, respectively. The pockets are nearly of the same size which results in the nesting properties of the electron and hole bands at the antiferromagnetic wave vector, ${\bf Q}_{AF}$, {\i.e.} $\varepsilon_{\bf k}^{e}=-\varepsilon_{\bf k+Q_{AF}}^{h}$.  Despite the fact that there are two electron and two hole pockets it has been argued\cite{Chubukov08} that it is enough to consider only two of them (one electron and one hole pocket) because the instabilities of the two-band model are the same as in the four-band model. Following this suggestion we consider
two bands that are given by diagonalized tight binding expression including hoppings up to the next nearest neighbors:
\begin{eqnarray}
{\cal H}
&=&
\sum\limits_{{\bf k}\gamma\sigma }\varepsilon _{{\bf k}\gamma}c_{{\bf k}\gamma\sigma
}^{\dagger}c_{{\bf k}\gamma\sigma }
+\sum\limits_{{\bf k}\gamma}\Delta_{{\bf k}}\left(
 c_{{\bf k}\gamma\uparrow }^{\dagger}c_{-{\bf k}\gamma\downarrow }^{\dagger}+h.c.\right)
\nonumber \\
&&
+\epsilon_f\sum\limits_\sigma f_\sigma ^{\dagger}f_\sigma
+\sum\limits_{{\bf k}\gamma\sigma }V_{{\bf k}\gamma}\left( c_{{\bf k}\gamma\sigma }^{\dagger}f_\sigma
+h.c.\right)
 \nonumber \\&&
+Uf_{\uparrow }^{\dagger}f_{\uparrow }f_{\downarrow}^{\dagger}f_{\downarrow }\quad,
\label{eq-01}
\end{eqnarray}
where $c_{{\bf k}\gamma\sigma
}^{\dagger}$ creates an electron with spin $\sigma$
in band  $\gamma$ ($\gamma=1,2$ refer to the electron and hole band, respectively)
with wave vector ${\bf k}=(k_x,k_y)$.
The dispersion $\varepsilon_{{\bf k}\gamma}$  is then given in the tight-binding form\cite{Korshunov08}
\begin{eqnarray}
\varepsilon_{{\bf k}1}&=& -0.18+ 0.16\left(\cos k_x+\cos k_y \right)-0.052  \cos k_x \cos k_y 
\nonumber\\
\varepsilon_{{\bf k}2}&=& 0.68+0.38\left(\cos k_x+\cos k_y \right)-0.8 \cos \frac {k_x}{2}\cos \frac{k_y}{2}.
 \nonumber\\&&
\label{eq-02}
\end{eqnarray}
Here, $\varepsilon_{{\bf k}1}$ dispersion yields the hole Fermi surface pocket around the $\Gamma$ point and $\varepsilon_{{\bf k}2}$ gives the elliptic electron Fermi surface pocket around the $M=(\pi,\pi)$ point, see Fig.\ref{figFS}. The parameters have been chosen from the available fit
to the ARPES data \cite{Nakayama09} (all in eV) and correspond to the hole doping of about 10$\%$.
The $f_\sigma ^{\dagger}$ operator create the localized electron at the impurity site at the origin
and $U$ is its on-site Coulomb repulsion.
 Finally $\epsilon_f$ is the f-band position, $V_{{\bf  k}\gamma}$ is the hybridization energy between localized electron
and the conduction bands, and $\Delta_{{\bf k}}$ is the singlet superconducting gap function.
We choose  values for $\epsilon_f$, and $V_{{\bf  k}\gamma}$ (Fig.\ref{Pspm1}) 
such that the f-orbital is almost filled ($n_f=1$).

Our model assumes the limit $U\rightarrow \infty$ where doubly occupied f-states are excluded.
This limit may be represented  by introducing the auxiliary boson $b$,  with the  constraint
$\tilde{Q}=b^{\dagger}b+\sum_{\sigma}f^{\dagger}_{\sigma}f_{\sigma}=1$\cite{Col84}. In the mean field approximation
($b=\langle b \rangle=\langle b^{\dagger} \rangle$),
the total Hamiltonian including  the constraint is given by ${\cal H}_{MF}+\lambda(b^2-1)$.
Here ${\cal H}_{MF}$ is obtained as
\begin{eqnarray}
{\cal H}_{MF} =\widehat{\varphi }^{\dagger}\beta_0\widehat{\varphi }+
\sum\limits_{{\bf k}}
\hat{\psi}_{{\bf k}}^{\dagger}\beta_1({\bf k})
\hat{\psi}_{{\bf k}%
}
+(
\hat{\psi}_{{\bf k}}^{\dagger}\beta_2
 \widehat{\varphi}
+h.c.),
\label{eq-03}
\end{eqnarray}
where $\lambda$ is  the Lagrange multiplier for enforcing the constraint. The Nambu
spinors  are denoted by
%
\begin{figure}
\centerline{
\includegraphics[angle=0,width=0.5\linewidth]{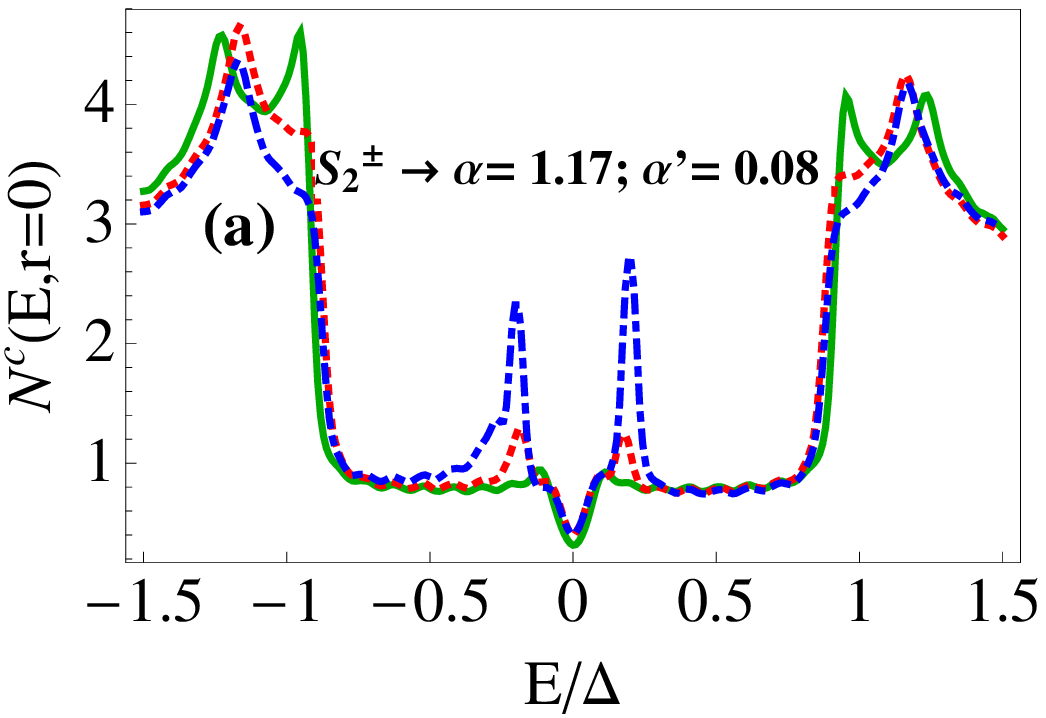}
\includegraphics[angle=0,width=0.5\linewidth]{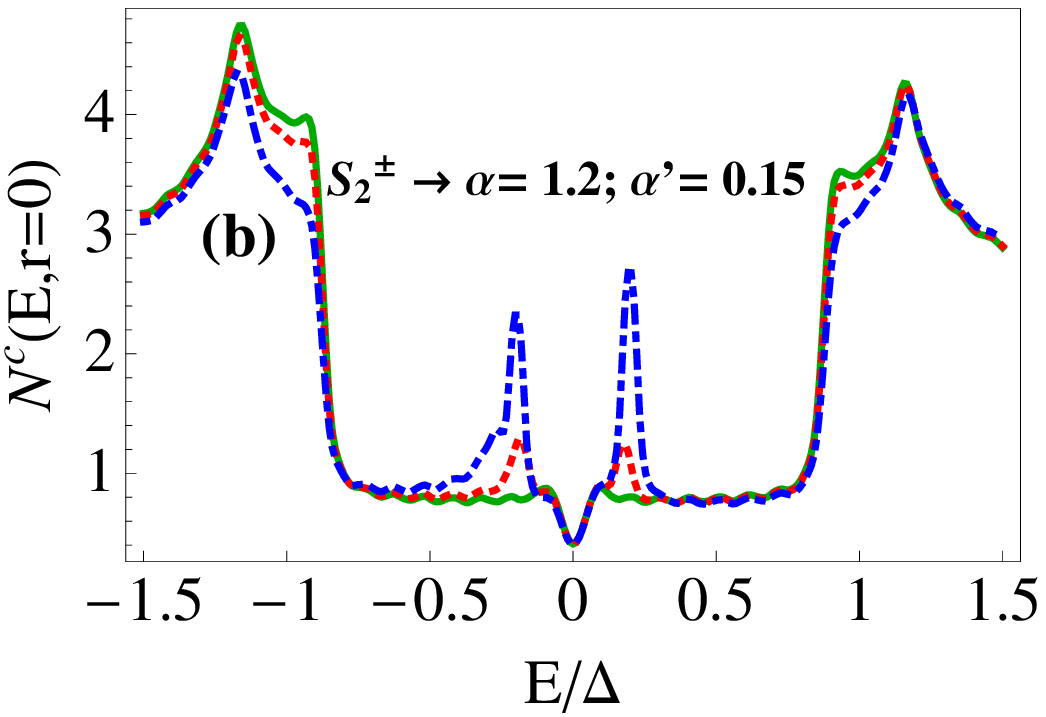}
}
\caption{(color online) LDOS for the superconducting state with various order parameters:  $\Delta_{S_2^\pm}(k_x,k_y)$ for
($\alpha=1.2;\; \alpha'=0.15$)
and ($\alpha=1.17;\; \alpha'=0.08$); for $\epsilon_f=-\Delta_0/3$, $V_{1}=V_{2}=0$ (green),
$V_{1}=V_{2}=0.5\Delta_0$ (red), and  $V_{1}=V_{2}=\Delta_0$ (blue).}
\label{Pspm2}
\end{figure}

$
\hat{\psi}_{{\bf k}}^{\dagger}=(
c_{{\bf k}1\uparrow }^{\dagger},
c_{-{\bf k}1\downarrow },c_{{\bf k}2\uparrow }^{\dagger},
c_{-{\bf k}2\downarrow })$, and likewise  $\qquad \widehat{\varphi }^{\dagger}=(
f_{\uparrow }^{\dagger},
f_{\downarrow },f_{\uparrow }^{\dagger},
f_\downarrow )
$,
while the matrices
$\beta_i$ are  defined as
\begin{eqnarray}
&&\beta_0={\tilde\epsilon_f}\varsigma_0\otimes\sigma_z;
\nonumber \\&&
\beta_1({\bf k})=\left[
\frac{1+\varsigma_3}{2}\otimes
\left( \varepsilon _{{\bf k}1}\sigma _z+\Delta _{{\bf k}}\sigma _x\right)\right.+
\nonumber \\&&\hspace{2cm}
\left.
\frac{1-\varsigma_3}{2}\otimes
\left( \varepsilon _{{\bf k}2}\sigma _z+\Delta _{{\bf k}}\sigma _x\right)
\right];
\nonumber \\&&
\hspace{0.5cm}\beta_2({\bf k})=\left[{\tilde V_{{\bf k}1}}\frac{1+\varsigma_3}{2}
+
{\tilde V_{{\bf k}2}}\frac{1-\varsigma_3}{2}
\right]\otimes
\sigma _z.
\label{eq-04}
\end{eqnarray}
Here $\sigma_i$
are the Pauli matrices acting in spin space, $\varsigma_i$ are the Pauli
matrices in the orbital space, and $\varsigma_i\otimes\sigma_j$ denotes a
direct product of the matrices operating on the 4-dimensional Nambu
space. Furthermore ${\tilde\epsilon_f}=(\epsilon_f+\lambda)/2$,
and $ \bar{V}_{{\bf k}\gamma}=bV_{{\bf k}\gamma}$ are effective hybridization
and energy position of the impurity $f$-level, respectively.

The local density of states  (LDOS) near the
magnetic impurity is obtained from analytic continuation $i\omega_n\rightarrow E+i 0^{+}$ according to
\bea
N^{c}(E,{\bf r})&=&-\frac{1}{\pi}
Im\left[ G^{c}_{11}({\bf r},{\bf r},\omega_n) 
+ G^{c}_{33}({\bf r},{\bf r},\omega_n)\right],
\label{eq-05}
\eea
where  $\omega_n=\pi T (2n+1)$ is the Matsubara frequency, and
 $
 G^{c}(r,r^\prime,\omega_n)$ is a Fourier transformation of
the matrix of the conduction electrons Green's function.

The matrix Green's functions are defined as the imaginary-time ordered
average
\bea\nonumber
 G^{c} ({\bf k}, {\bf k}^\prime;\tau)&=&
-\langle T_{\tau} \hat{\psi}_{{\bf k}}(\tau) \hat{\psi}_{{\bf k\prime}}^{\dagger}(0)\rangle;
\\ \nonumber
G^{fc} ({\bf k};\tau)&=&
-\langle T_{\tau}  \widehat{\varphi }(\tau)\hat{\psi}_{{\bf k}}^{\dagger}(0)\rangle;
\\\nonumber
 G^{cf} ({\bf k},\tau)&=&
-\langle T_{\tau} \hat{\psi}_{{\bf k}}(\tau) \widehat{\varphi }^{\dagger}(0)\rangle;
\\
 G^{f} (\tau)&=&
-\langle T_{\tau} \widehat{\varphi }(\tau)  \widehat{\varphi }^{\dagger}(0)\rangle.
\label{eq-06}
\eea
Where $G(\tau)=T\sum_{\omega_n}G(\omega_n)e^{-i\omega_n\tau}$. At  low temperature regime  $T\sum_{\omega_n} (\ldots)\longrightarrow \frac{-1}{\pi}Im\int_{0}^{D_{\gamma}}d\omega\lim\limits_{i\omega\rightarrow \omega+i0^{+}}(\ldots)$, where $D_{\gamma}$ is bandwidth of conduction band $\gamma$. 
Using the standard equations of motion method, one can show that
\bea
 (i\omega_n-\beta_1({\bf k}))  G^{c} ({\bf k}, {\bf k}^\prime,\omega_n)=\delta_{{\bf k}, {\bf k}^\prime}
+\beta_2({\bf k})G^{fc} ({\bf k}^\prime,\omega_n);\nonumber\\
\label{eq-07}
\eea
\bea
 (i\omega_n-\beta_0)
 G^{fc} ({\bf k},\omega_n)=
\sum\limits_{{\bf k}'}\beta_2({\bf k}')G^{c} ({\bf k}',{\bf k},\omega_n);\label{Gfc}
%
\label{eq-08}
\eea
and
\bea
 (i\omega_n-\beta_0) G^{f} (\omega_n)=1+\sum\limits_{{\bf k}}\beta_2({\bf k})G^{cf} ({\bf k},\omega_n);
\label{eq-09}
\eea
\bea
 (i\omega_n-\beta_1({\bf k}))G^{cf} ({\bf k},\omega_n)=\beta_2({\bf k})G^{f} (\omega_n).
\label{eq-10}
\eea
Now using the equations above we find the full $f$-Green's function
\bea
 G^{f} (\omega_n)= \frac{1}{(i\omega_n-\beta_0-\Sigma_f)},
\label{eq-11}
\eea
where the $f$-self energy is given by
\bea
\Sigma_f&=&\sum\limits_{{\bf k}}\beta_2({\bf k}) G_{0}^{c}({\bf k},\omega_n) \beta_2({\bf k})
\nonumber\\
&=& \frac{1+\varsigma_3}{2}\otimes
\sum\limits_{{\bf k}}{\tilde V}_{1{\bf k}}^{2}\left(\frac{-i\omega_n- \epsilon_{{\bf k}1}\sigma_z+
\Delta_{{\bf k}1}\sigma _x}{\omega_n^2+ \epsilon_{{\bf k}1}^2+ \Delta_{{\bf k}1}^2}
\right)
\nonumber\\&&
+(1\rightarrow 2;\;\;\varsigma_3\rightarrow -\varsigma_3),
\label{eq-12}
\eea
and 
the conduction electrons Green's function can be obtained by\cite{Zhang01} 

\bea
 G^{c} ({\bf k}, {\bf k}^\prime,\omega_n)=G_{0}^{c}({\bf k},\omega_n)[\delta_{{\bf k}, {\bf k}^\prime} +
t({\bf k},{\bf k^\prime};\omega_n)G_{0}^{c}({\bf k}^\prime,\omega_n)].
\nonumber\\
\label{eq-13}
\eea
Here, $G_{0}^{c}({\bf k},\omega_n)=\left( i\omega_n-\beta_1({\bf k}) \right)^{-1}$ is the unperturbed Green's
function of the conduction electrons, and  the $t$-matrix is given by
\be 
t({\bf k},{\bf k^\prime};\omega_n)=\beta_2({\bf k}) G^{f} (\omega_n)\beta_2({\bf k^\prime}).
\label{eq-14}
\ee

In the following, the ${\bf  k}$ dependence of the hybridization energy is neglected, {\it i.e.}, we set
$V_{{\bf  k}\gamma}=V_{\gamma}$, which yields that $\beta_2({\bf  k})\rightarrow \beta_2$.

Minimization of the ground state energy with respect to
 $b$  and the Lagrange multiplier $\lambda$ leads to the mean field equations
\bea
\lambda b^2=\sum\limits_{{\bf k}\gamma\sigma}{\tilde V}_{{\bf k}\gamma}W^{fc}_{{\bf k}\gamma\sigma},
\hspace{1cm} \sum\limits_{\sigma}n_{\sigma}^{f}+b^2=1,
\label{MF-eq}
\label{eq-15}
\eea
where the expectation values are defined by
$W^{fc}_{{\bf k}\gamma\sigma}=\langle f^\dagger_{\gamma}c_{{\bf k}\gamma\sigma
} \rangle$,
and $n_{\sigma}^{f}=\langle f_{\sigma}^\dagger f_{\sigma}\rangle$.
Therefore  from Eq.~(\ref{MF-eq}) we show easily that
\bea
\lambda b^2&=&\lim\limits_{\tau\longrightarrow 0}\sum\limits_{{\bf k}}
\left[{\tilde V}_{1} \left(G^{fc}_{11} ({\bf k},\tau) -  G^{fc}_{22} ({\bf k},\tau)\right) \right.
\nonumber\\&&
\left.+{\tilde V}_{2}\left(G^{fc}_{33} ({\bf k},\tau) -  G^{fc}_{44} ({\bf k},\tau)
\right)\right],
\label{eq-16}
\eea
and
\bea 
b^2=
\frac{1}{2}\lim\limits_{\tau\longrightarrow 0}\left[G^{f}_{11} (\tau)-G^{f}_{22} (\tau)
\right.
\left.
+ G^{f}_{33} (\tau)-G^{f}_{44} (\tau)\right].
\label{eq-17}
\eea
By solving the set of equations~(\ref{eq-11})-(\ref{eq-17}), numerically 
one can find the values of $\tilde{\epsilon}_f$ and $b$ 
which are used as an input for the $t$-matrix..

\begin{figure}
\centerline{
\includegraphics[angle=0,width=0.5\linewidth]{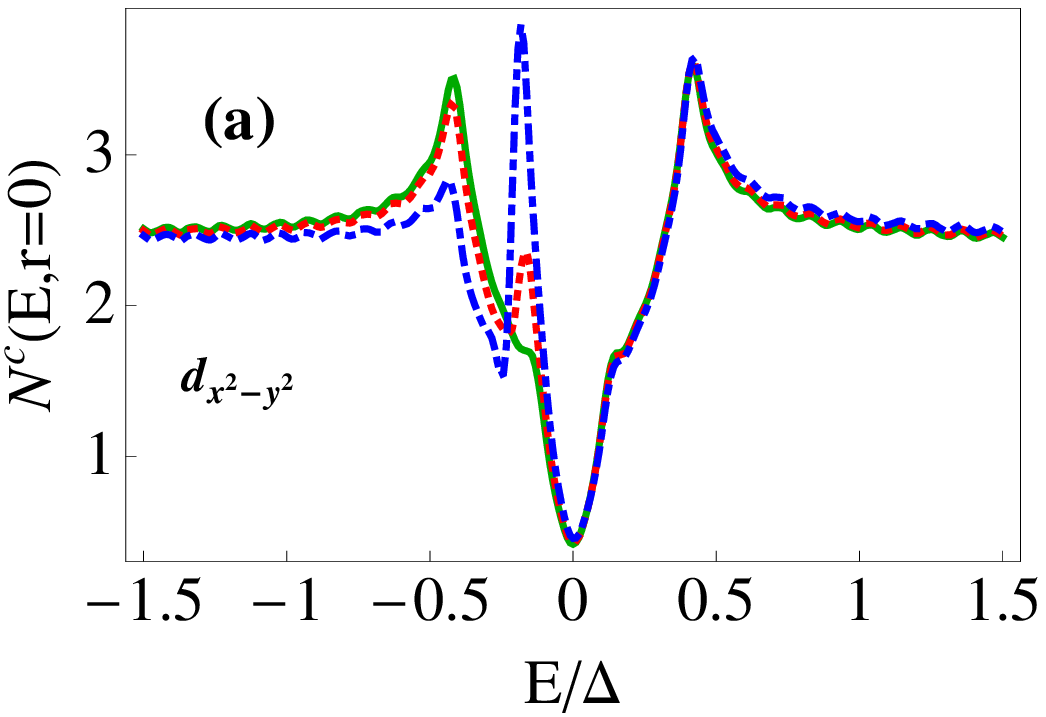}
\includegraphics[angle=0,width=0.5\linewidth]{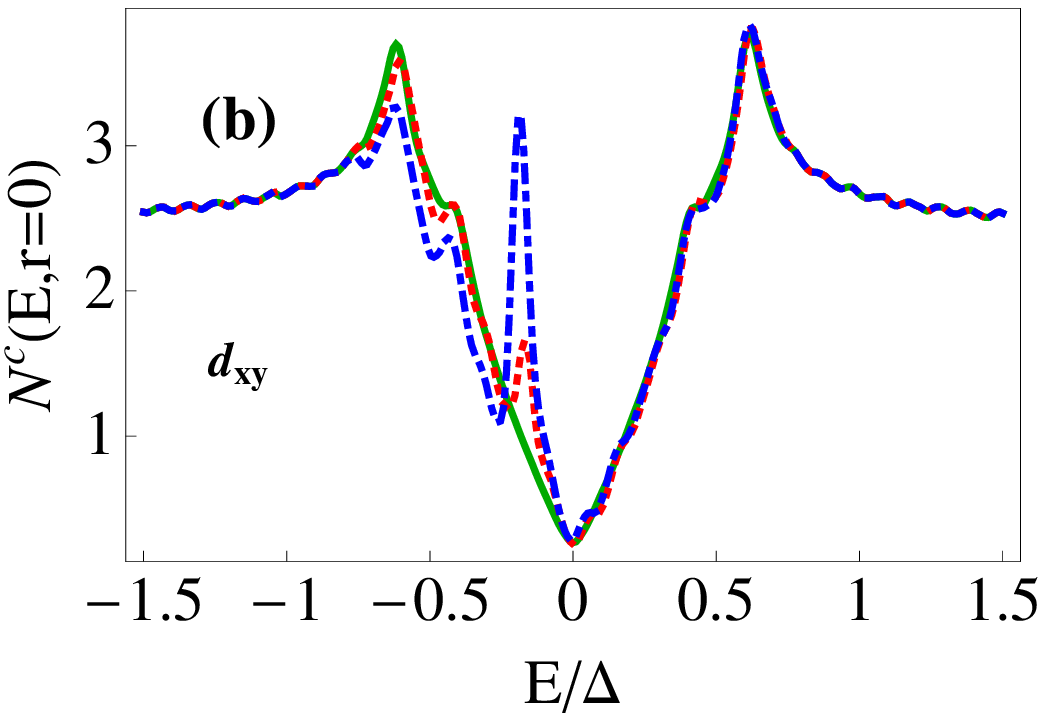}
}
\caption{(color online) LDOS for the superconducting 
regime with order parameters: a) d$_{x^2-y^2}$
 and b) d$_{xy}$; for $\epsilon_f=-\Delta_0/3$,
$V_{1}=V_{2}=0$ (green), $V_{1}=V_{2}=0.5\Delta_0$ (red),
and  $V_{1}=V_{2}=\Delta_0$ (blue).}
\label{Pd}
\end{figure}


\section{Numerical results}
\label{sect:numerics}

We now discuss the results of numerical calculations for the central quantity 
$N^c(E,{\bf r})$ based on the previous analysis (obtained from Eqs.~\ref{eq-11}-\ref{eq-17}).
In this section we focus on  the energy and also spatial dependence of the local density of states (LDOS) (Eq.~\ref{eq-05}) at $T = 0$ for various gap symmetries.
As the main candidates we include different types of the extended s-wave gaps which are fully gapped on the hole pocket but possibly have accidental nodes on the electron pockets due to higher harmonics contributions. For completeness, we also include d-wave gap functions
which have nodes on both electron and hole pockets. The latter however are not supported by ARPES results \cite{Ding08,Kondo08,Lue09} which suggest nodeless gaps on the hole pockets.

We begin our discussion with the anisotropic nodeless extended s-wave pairing function (S$^\pm$).
The Fermi surface  illustration with respective node positions indicated by dashed lines is shown in Fig.\ref{figFS}.(a).
This gap function is given by
\bea
\Delta_{S^\pm}(k_x,k_y)=
\frac{\Delta_0}{2} (\cos  k_x+\cos   k_y),\eea
and we show the result for the LDOS at the origin $(r_x,r_y)=(0,0)$ in Fig.\ref{Pspm1}(a) for $\Delta_0=6$ meV
and  $\epsilon_f=-\Delta_0/3$, with different hybridization energies
$V_{1}=V_{2}=0$ (green), $V_{1}=V_{2}=0.5\Delta_0$ (red),
and  $V_{1}= V_{2}=\Delta_0$ (blue). While the overall structure of the spectrum stays the same
the increased hybridization leads to more pronounced bound state peaks within the gap.
 We observe 
that an increase of the hybridization energy, $V_{\gamma}$, causes the
position of peaks to move to  higher absolute values of $E$, 
and a corresponding increase of their line width.
By increasing  the absolute value of the impurity energy level, $\epsilon_f$ , the bound state
also moves to  higher absolute values of $E$.
Notice that by restricting to
the first order perturbation theory in effective hybridization 
(Born approximation), we did not find a 
dramatic change in our results, for the chosen values of $V_{\gamma}$.

Furthermore, due to the different size of electron and hole pockets the onset of the continuum around $|E/\Delta|$ is split into
a double peak structure. For $E>0$ the lower and upper peaks correspond to hole and electron pockets, respectively.

The addition of higher harmonics in the S$_1^\pm$ gap function allows to tune the modulus of the gaps on hole and
electron pockets independently while keeping the basic property of having an opposite signs. As seen from Fig.\ref{figFS}(a) this
is equivalent to shift of the nodal line position closer to the electron pockets and increasing higher harmonic amplitude continuously one
eventually produce an accidental node on this sheet. This gap function is given by
\bea
&&\Delta_{S_1^\pm}(k_x,k_y)=\nonumber \\&&
\Delta_0\left[\frac{1}{2}(\cos  k_x+\cos   k_y)
+\alpha \cos \frac{k_x}{2} \cos \frac{k_y}{2}\right],
\eea
and the result for the LDOS is plotted in Fig.(\ref{Pspm1}(b)-(f) for different $\alpha$ parameters.
At the symmetry points one has $\Delta_{(0,0)}=\Delta_0(1+\alpha)$
and $\Delta_{(\pi,\pi))}=-\Delta_0$. Then $|\Delta_{(0,0)}|-|\Delta_{(\pi,\pi))}|=\alpha$ which also means that the difference between
the absolute values of
the gaps on $\Gamma$-centered hole pockets and M-centered electron pockets increases with $\alpha$. This can be clearly seen in
Fig.~\ref{Pspm1}(b) ($\alpha = 1$) where the peak at larger $|E/\Delta_0|$ originating from the hole pocket is pushed to larger energies
and for $\alpha=3$  is no longer visible on the scale of Fig.~\ref{Pspm1}(c). On the other hand the peak due to gap maximum on the
electron pocket stays fixed around $|E/ \Delta_0|\simeq 1$ while $\alpha$ grows. At the same time a deep minimum and finally an accidental node of the gap develops on the electron pocket and leads to the increase of the low energy LDOS in  Figs.\ref{Pspm1}(b)-(f). This type of low energy DOS may explain  power laws for NMR relaxation rate and penetration depth observed in some pnictides\cite{Nakai08,Shan08,Mu08}.
On the other hand the position of bound state peaks caused by the magnetic impurity is apparently insensitive to the variation of $\alpha$ and the change of the underlying quasiparticle spectrum.

However, the S$_1^\pm$  gap function is not unique and accidental nodes on the electron pocket may be obtained with a different
type of modification with higher harmonics. This leads us to extended  S$_2^\pm$-wave pairing described by another gap function
\bea
&&\Delta_{S_2^\pm}(k_x,k_y)=\Delta_0
\left[0.5(\cos k_x+\cos k_y) \right.
\nonumber\\&&
\left.+\alpha \cos  k_x \cos k_y
+\alpha' \cos 4 k_x \cos 4 k_y\right].
\eea
Its nodal structure is shown in (Fig.~\ref{figFS}(b). Now there are two nodal lines, one located between the pockets
which leads to an anisotropic but fully gapped order parameter on the hole pockets. The other accidental nodal line centered around
the M point cuts the electron pocket and leads to a finite low energy quasiparticle DOS as seen in the results of
the Figs.(\ref{Pspm2}(a) and \ref{Pspm2}(b) for  $\alpha=1.2;\; \alpha'=0.15$ and $\alpha=1.17;\; \alpha'=0.08$ respectively.
The bound states due to impurity scattering appear again as pairs at similar energies as for the S$_1^\pm$ case.

Finally for completeness we also consider two simple anisotropic d-wave order parameters, namely d$_{x^2-y^2}$  (Fig.~\ref{figFS}(d)) and
d$_{xy}$  (Fig.~\ref{figFS}(e)) which have symmetry enforced gap nodes. This leads to
  sign change of the gap function on the same FS pocket rather than between them. We consider the two candidates
\bea
\Delta_{d_{x^2-y^2}}(k_x,k_y)&=&
\frac{\Delta_0}{2} \left(\cos  k_x-\cos   k_y\right),\nonumber\\
\Delta_{d_{xy}}(k_x,k_y)&=&\Delta_0
\sin  k_x\sin   k_y.
\eea
In each case the nodal lines cross {\it both} Fermi surface pockets in contrast to the extended s-wave model.
We note that current ARPES experiments have shown a fully gapped
hole pocket \cite{Ding08,Kondo08,Lue09} though no experiments yet are available for $P$-based systems.

The background of the LDOS is given by the typical V-shape of a d-wave order parameter. On top of it
a single bound state peak due to the impurity scattering appears below the Fermi level. This is distinctly
different from the extended s-wave case where always two bound states below and above the Fermi level
appear symmetrically. A partly similar observation for a d$_{x^2-y^2}$ order parameter with only a single
sheet FS intended for cuprates was made in Ref.~\onlinecite{Zhang01}. There one bound state peak was found
for $(r_x,r_y)$ along the anti-nodal direction and two peaks for the nodal direction. In our present d-wave
case considered for the two-sheet FS of Fe pnictides the single peak appears for both nodal and anti-nodal
directions.

\begin{figure}
\centerline{\includegraphics[angle=0,width=0.5\linewidth]{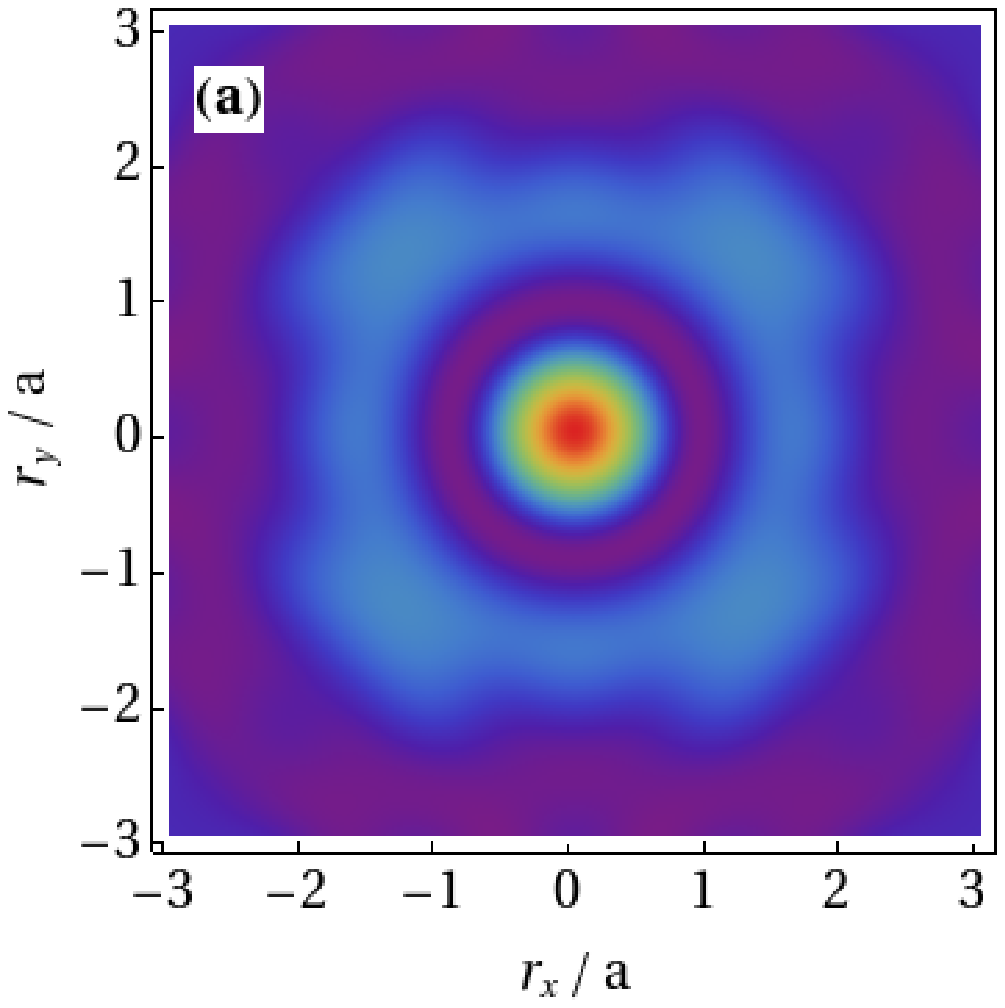}
\includegraphics[angle=0,width=0.5\linewidth]{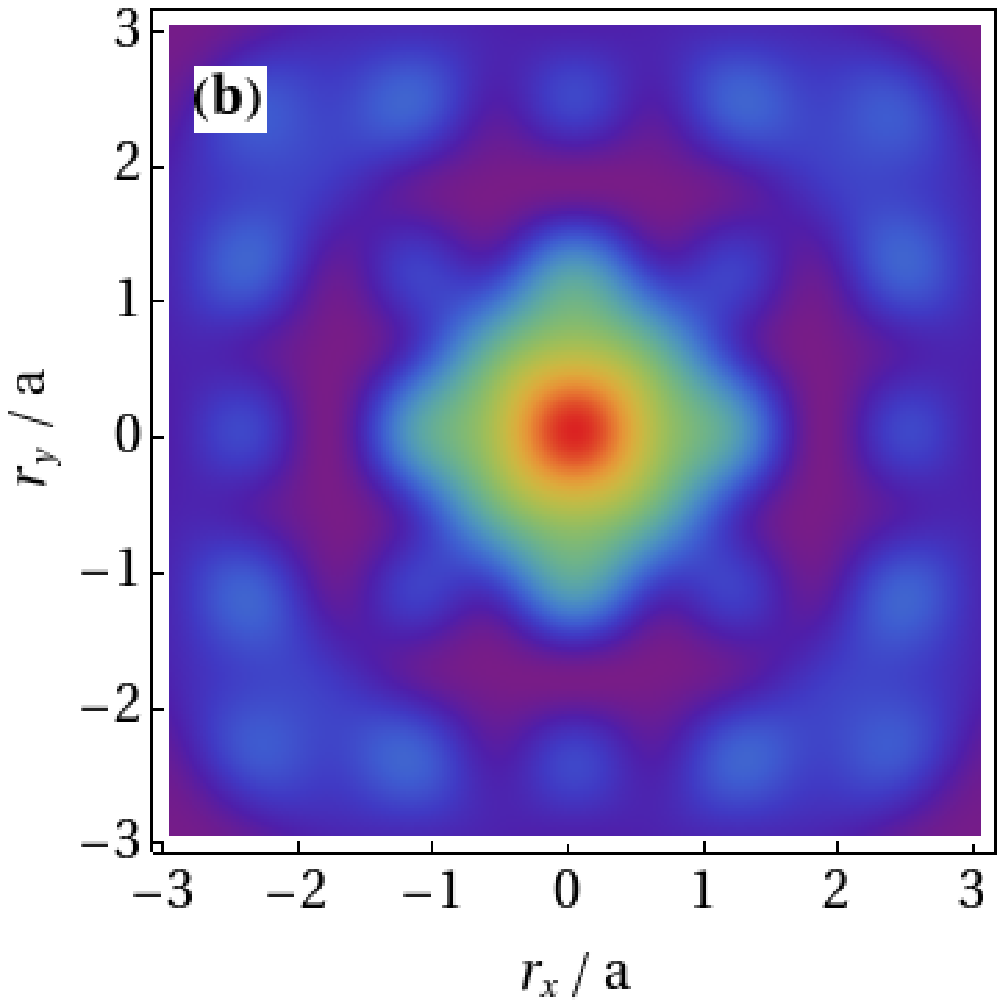}}
\caption{(color online) Density plot of the spatial distribution of the LDOS for
the superconducting regime with extended s-wave order parameter,  $\Delta_{S^\pm}(k_x,k_y)$,
for $\epsilon_f=-\Delta_0/3$,  $V_{1}=V_{2}=\Delta_0$, and a) $\omega=\omega_r$,  b) $\omega=-\omega_r$. (Red color correspond to large LDOS).}
\label{Pspm_mesh}
\end{figure}

For clarifying of the position dependence of resonance peaks,  Fig.\ref{Pspm_mesh}
displays the spatial variation of the LDOS $N_c(\omega,r_x,r_y)$  around the magnetic impurity at the resonance energies,
(a) $\omega=\omega_r$ and (b) $\omega=-\omega_r$,
for  S$^\pm$ gap symmetry with $\Delta_0=6$meV;
 $\epsilon_f=-\Delta_0/3$, and $V_{1}=V_{2}=\Delta_0$.
It shows that the maximum amplitude of the LDOS appears
close to the impurity site and decays non-monotonically further away from the impurity site. While
the LDOS at $\omega_r$ is rather isotropic the peak for $-\omega_r$ shows a significant anisotropic
LDOS in the plane. This anisotropy does not seem to result from special FeAs Fermi surface feature
since it is also observed in the single parabolic band case in Ref.~\onlinecite{Zhang01}.

\section{Summary}
\label{sect:summary}

We have investigated the effect of magnetic impurity scattering in the FeAs pnictide superconductors. We used a simple two band
model Fermi surface and calculated LDOS spectral and spatial dependence close to the impurity site for two types of extended s-wave
superconducting order parameters with inter-band sign change and for two d-wave order parameters with intra-band sign change. In
the former two impurity bound states appear symmetrically around the Fermi energy at positions $\pm\omega_r$. The modulus of the bound state
energy increases with hybridization strength  $V$  and impurity orbital energy $\epsilon_f$ monotonically.
In the latter case only the lower bound state pole at $-\omega_r$ appears in the LDOS for any direction from the impurity site.
The background variation of the LDOS is determined by the characteristics of the superconducting gap on the two FS sheets. The extended S-wave order
parameters may be tuned such that fully gapped behavior on the central hole sheet and accidental node structure on the zone boundary hole sheets
appear naturally. In this case the spatial dependence of the LDOS for the two bound state peaks shows significant differences in the degree of spatial anisotropy.
We conclude that the observation of two bound state peaks in tunneling experiments would be an important support for the extended s-wave gap function
with interband sign change. The fine structure of the background continuum LDOS may give more detailed information on the type of the accidental nodal structure.


\end{document}